\documentclass[a4paper,11pt,amssymb,eqsecnum]{article}
\pdfoutput=1 % if your are submitting a pdflatex (i.e. if you have
\usepackage{jcappub} % for details on the use of the package, please
\usepackage{graphicx}
\usepackage{amsmath}
\usepackage{amssymb}
\usepackage{amsfonts}
\usepackage{url}
\usepackage{color}
\usepackage{mathtools}
\newcommand{\be}{\begin{equation}}
\newcommand{\ee}{\end{equation}}

\def\bea{\begin{eqnarray}}
\def\eea{\end{eqnarray}}
\def\lcdm{$\Lambda$CDM }
\newcommand{\ba}{\begin{eqnarray}}
\newcommand{\ea}{\end{eqnarray}}

\title{\boldmath Machine Learning meets the redshift evolution of the CMB Temperature}

\author[a]{Rub\'{e}n Arjona,\note{Corresponding author.}}

\affiliation[a]{Instituto de F\'isica Te\'orica UAM-CSIC, Universidad Auton\'oma de Madrid,
Cantoblanco, 28049 Madrid, Spain}

\emailAdd{ruben.arjona@uam.es}

\abstract{We present a model independent and non-parametric reconstruction with a Machine Learning algorithm of the redshift evolution of the Cosmic Microwave Background (CMB) temperature from a wide redshift range $z\in \left[0,3\right]$ without assuming any dark energy model, an adiabatic universe or photon number conservation. In particular we use the genetic algorithms which avoid the dependency on an initial prior or a cosmological fiducial model. Through our reconstruction we constrain new physics at late times. We provide novel and updated estimates on the $\beta$ parameter from the parametrisation $\text{T}(z)=\text{T}_0(1+z)^{1-\beta}$, the duality relation $\eta(z)$ and the cosmic opacity parameter $\tau(z)$. Furthermore we place constraints on a temporal varying fine structure constant $\alpha$, which would have signatures in a broad spectrum of physical phenomena such as the CMB anisotropies. Overall we find no evidence of deviations within the $1\sigma$ region from the well established $\Lambda\text{CDM}$ model, thus confirming its predictive potential.}

\begin{document}
\maketitle
\flushbottom

\section{Introduction}\label{section:introduction}
Our current knowledge for the evolution of the Universe as a whole from the first fraction of a second to our present day, about $13.6$ billion years later, rests upon the successful hot Big Bang cosmological model \cite{GarciaBellido:2005df}. It is built on the robust theoretical framework of General Relativity (GR) and based on well tested observations such as the expansion of the Universe \cite{Hubble:1929ig}, the relative abundance of light elements \cite{Gamow:1946eb} and the cosmic microwave background (CMB) \cite{Penzias:1965wn}. The presence of the latter is considered to be the best indication for a primordial expanding state of the Universe originating from an initial high density state to become an almost perfect isotropic blackbody radiation at a temperature of about 3K and whose emission we receive around $380,000$ years after the Big Bang.\\
The hot Big Bang model predicts that the CMB photon energy is redshifted with the cosmic expansion. In other words, the Universe has a hot and dense past and cools as it expands adiabatically according to the linear average temperature-redshift relation (TRR) of the CMB, $\text{T}_{\text{CMB}}(z)=\text{T}_0(1+z)$ where $\text{T}_0=(2.72548\pm0.00057)\text{K}$ is the local measurement of the CMB temperature today i.e. at $z=0$ \cite{Fixsen:2009ug} and $\text{T}_{\text{CMB}}(z)$ represents the temperature measured by an observer at redshift $z$. This relation is not confined to a specific metric theory, holding in the framework of GR and the electromagnetic theory of Maxwell under the assumption that photons are massless, the CMB is thermal radiation, the first law of thermodynamics is true and that the expansion of space is isotropic \cite{Abitbol:2019ewx}. Although this linear temperature relation is well established \cite{Srianand:2000wu} and departures from it would require important distortions in the Planck spectrum of the CMB \cite{Chluba:2014wda}, it can be modified for example \cite{Lima:2000ay} by adding extra components such as a decaying vacuum energy density or some process of quantum gravitational origin that could affect the adiabatic photon production (or destruction), due to late inflationary models induced by a scalar field, in string theory models where axions and photons could be mixed \cite{Jaeckel:2010ni} or theories with deviations from homogeneity and isotropy.\\
To constrain deviations from adiabatic evolution, the following power-law $\text{T}_{\text{CMB}}(z)=\text{T}_0(1+z)^{1-\beta}$  is commonly used, where $\beta$ is the parameter of the theory and $\beta>0$ or $\beta<0$  would be compatible with net photon production or destruction respectively \cite{Chluba:2014wda}. It is also of great interest models where some fundamental constants are not space-time invariant, such as the fine structure constant $\alpha$. This effect can be found for example in theories with extra compact dimensions (aiming to unify gravity and other fundamental forces), where the cosmic evolution of the scale factor will have a time dependence on the coupling constants \cite{Damour:1994zq}. A different possibility is the inclusion of a new scalar field with couplings to the Maxwell scale factor $F_{ab}F^{ab}$ whose evolution involves a variation of $\alpha$ \cite{deMartino:2016bjx}.\\
The Standard Cosmological Model, which contains a cosmological constant $\Lambda$, Cold Dark Matter (CDM) and is built on the hot Big Bang Theory and the Inflationary Paradigm, has become the best phenomenological description for the current accelerating expansion of the Universe. Yet there exists also plenty of other models that could account for the accelerated expanding Universe without the need of a cosmological constant $\Lambda$, whose nature still remains unknown, for example through the inclusion of scalar fields in Dark Energy (DE) models or through Modified Gravity (MG) theories. Even though DE and MG theories are driven by different physical backgrounds,  it has been shown that both kind of models can be studied on the same ground in an effective fluid approach \cite{Arjona:2019rfn,Arjona:2018jhh}. One of the first convincing evidences for a cosmic expansion came in 1998 as a result of an unanticipated dimming through the observed light of type Ia supernovae (SNIa) \cite{riess1998observational}. Although the cosmic acceleration has been asserted through other independent probes like the CMB \cite{Sherwin:2011gv}, baryon acoustic oscillations (BAO) \cite{Blake:2012pj} or the Hubble parameter \cite{Aghanim:2018eyx}, the presence of some cosmic opacity that can contribute to astronomical photometric measurements of distant SNe Ia is still an open possibility \cite{Holanda:2018oaj}. As mentioned in Ref.~\cite{Zhou:2019gda}, opacity sources could come from the non-conservation of the photon number density, which in turns changes the temperature-redshift and the distance duality relation, or from MG theories with non-minimal couplings between the electromagnetic Lagrangian and a new scalar field \cite{Holanda:2016wut}. If there is some extra dimming contribution, this would have an imprint in the cosmological parameters and the expansion rate inferred from SNe Ia measurements. Hence, testing the cosmic opacity parameter denoted as $\tau(z)$ and in turn the duality relation $\eta(z)$ where both are related through the parametrization $e^{\tau(z)/2}=\eta(z)$ \cite{Zhou:2019gda} is of great interest.\\
Machine Learning algorithms are successful at processing and extracting crucial information from large amounts of data and can remove the problem of model bias \cite{Ntampaka:2019udw}. They are also very useful to test the consistency of the dataset in a model independent approach and also to search for tensions or systematics. In this paper we will use a particular Machine Learning (ML) algorithm, the genetic algorithms (GA). The robustness of the GA resides in the fact that is a non-parametric method and does not require an initial prior or a cosmological fiducial model \cite{Bogdanos:2009ib,Nesseris:2012tt}. Even though the temperature-redshift relation (TRR) appears to be well established, measurements of the connection between the redshift and the CMB temperature serves as an important cosmological probe. Among others, it can help to confirm the assumption of photon number conservation, the CMB is thermal radiation, entropy conservation and that the expansion of space is isotropic. It provides also a way to discriminate alternative cosmologies \cite{lima2000radiation,lima1996frw}. Measurements of the TRR can be also used to measure the local expansion rate $H_0$ through the time evolution of the background $\text{T}_{\text{CMB}}(z)$ \cite{Abitbol:2019ewx}.\\
In this paper we implement the GA which is a model independent and non-parametric algorithm to reconstruct the evolution of the CMB temperature from a wide redshift range $z\in \left[0,3\right]$ without assuming any dark energy model, an adiabatic universe or photon number conservation. We then provide novel and updated estimates on the $\beta$ parameter from the parametrisation $\text{T}_{\text{CMB}}(z)=\text{T}_0(1+z)^{1-\beta}$, the duality relation $\eta(z)$ and the cosmic opacity $\tau(z)$. Furthermore we place constraints on a temporal varying fine structure constant $\alpha$, which could affect among others the CMB anisotropies \cite{Menegoni:2009rg,Uzan:2010pm}. We want to stress that our constraints are not independent of each other since all of them parameterize in diverse ways potential deviations from the temperature-redshift relation.\\
This paper is organized as follows. In Section~\ref{sec:analysis} we present the notation and methodology of our analysis with the minimal assumptions made for the reconstruction of the data. In Section~\ref{sec:results} we present our constraints and results and in Section~\ref{sec:conclusions} we present our Conclusions. In Appendix~\ref{sec:fisher} we compare our error analysis of the GA with the Fisher matrix approach and in Appendix~\ref{sec:errors} we describe the data used and our implementation for the error analysis.  Finally, in Appendix~\ref{sec:GA} we discuss in detail the theoretical background of the GA implementation.

\section{Analysis \label{sec:analysis}}
Genetic algorithms (GA), which are a particular ML method, implement adaptive heuristic search approaches based on grammatical evolutionary algorithms and inspired by biological evolution operations of mutation and crossover \cite{holland1992adaptation}. It has been proven to be remarkably useful and robust in a wide range of areas such as particle physics \cite{Abel:2018ekz,Allanach:2004my,Akrami:2009hp}, cosmology \cite{Bogdanos:2009ib,Nesseris:2010ep}, astronomy and astrophysics \cite{wahde2001determination,Rajpaul:2012wu} and other fields like computational science, economics, medicine and engineering \cite{affenzeller2009genetic,sivanandam2008genetic}. For further details on the GA and more applications to cosmology see \cite{Bogdanos:2009ib,Nesseris:2012tt}. For other successful symbolic regression methods applied in physics and cosmology see \cite{Udrescu:2019mnk,Setyawati:2019xzw,vaddireddy2019feature,Liao:2019qoc,Belgacem:2019zzu,Li:2019kdj,Bernardini:2019bmd,Gomez-Valent:2019lny}.\\
Mock datasets have already tested the reconstruction method of the GA approach \cite{Nesseris:2012tt,Nesseris:2014qca} and recently they have been used to reconstruct the deceleration parameter $q(z)$, which quantifies the acceleration of the Universe, making a $\sim4.5\sigma$ model independent confirmation of the accelerated expansion \cite{Arjona:2019fwb}. Moreover, through a unified GA analysis using data from the Hubble expansion $H(z)$, Baryon Acoustic Oscillations (BAO),the growth-rate and $Eg$ data \cite{Arjona:2020kco} there appears to be hints for the existence of an important non-adiabatic contribution to the dark energy (DE) sound speed or the presence of DE anisotropic stress, which if not considered can bias the cosmological parameters deduced from the data \cite{Cardona:2019qaz}, thus pointing out to possible deviations from the $\Lambda\text{CDM}$ model.\\
On the other hand, Gaussian Processes (GP), which likewise have been applied to cosmology \cite{Shafieloo:2012ht,Busti:2014aoa,Pinho:2018unz,Bengaly:2019oxx}, are also non-parametric methods for data reconstruction and assumes that the stochastic data is characterized by a Gaussian process that can be mapped to a cosmological function of concern. Although GP requires the choice of a kernel function and a fiducial model, normally  $\Lambda$CDM, it has been asserted that this doesn't have an effect in the reconstruction \cite{Shafieloo:2012ht}.\\
In this section we apply the GA to reconstruct the background temperature of the CMB $\text{T}_{\text{CMB}}(z)$ given in Table~\ref{tab:data}. Hereafter we will express $\text{T(z)}\equiv \text{T}_{\text{CMB}}(z)$. In our analysis we use 37 points of the compilation from Table~\ref{tab:data} which spans over a wide redshift range of $0\le z \le 3.025$. The data is in the form $\left(z_i, T_i,\sigma_{T_i}\right)$. Since our $\chi^2$ has a quadratic form
\begin{equation}
  \chi^2_{T}=\sum_i^{N_T}\left(\frac{T_i-T^{th}(z_i)}{\sigma_{T_i}}\right)^2,
\end{equation}
where $\text{T}^{\text{th}}(z)=\text{T}_0\tilde{\text{T}}(z)$ and $\tilde{\text{T}}(z)$ is the dimensionless temperature, we can minimize the $\chi^2$ analytically over $T_0$ finding
\ba\label{eq:To}
\chi^2_{T}&=&A-\frac{B^2}{\Gamma},\label{eq:chi2H}\\
T_0&=&\frac{B}{\Gamma},\label{eq:H0bf}
\ea
where the parameters $A$, $B$ and $\Gamma$ are defined as
\ba
A&=&\sum_i^{N_T}\left(\frac{T_i}{\sigma_{T_i}}\right)^2, \\
B&=&\sum_i^{N_T}\frac{T_i~\tilde{\text{T}}(z_i)}{\sigma_{T_i}^2}, \\ \Gamma&=&\sum_i^{N_T}\left(\frac{\tilde{\text{T}}(z_i)}{\sigma_{T_i}}\right)^2,
\ea
and we denote the theoretical value $\text{T}^{\text{th}}(z)$ of the background temperature of the CMB obtained from the GA as $\text{T}^{\text{th}}(z)=\text{T}^{GA}(z)$. Then $\tilde{\text{T}}(z)=\text{T}^{GA}(z)/\text{T}_0$ and we set $N_T=37$. Our best-fit function found is
\begin{equation}\label{eq:reconst}
  T^{GA}(z)=T_0\left(1+z\left(e^{0.00123z^2}-0.03581z+0.00678z^2\right)\right),
\end{equation}
and our own assumption is that the value of the background temperature of the CMB $\text{T(z)}$ today is given by $\text{T}(z=0)=\text{T}_0$ where $\text{T}_0$ is obtained directly from Eq.~(\ref{eq:H0bf}). The best-fit $\chi^2$ for the GA is $\chi^2=28.816$, which is smaller than that of the $\Lambda\text{CDM}$ model with a $\chi^2$ of $\chi^2=29.176$ or that with the Fisher matrix approach with $\chi^2=28.892$, see Appendix~\ref{sec:fisher}. In the left panel of Fig.~\ref{fig:Data} we present the $\text{T}(z)$ data compilation shown as grey points along with the \lcdm best-fit (dashed line) and the GA best-fit (solid black line). The shaded gray regions corresponds to the $1\sigma$ errors of the GA. For the evaluation of the GA errors on the reconstructed quantities we make use of the path integral approach, first derived in \cite{Nesseris:2012tt,Nesseris:2013bia}, where one calculates analytically a path integral over the functional space that can be scanned by the GA. We have also tested our GA approach leaving free the function $f_{GA}(z)$ as $\text{T}^{GA}(z)=\text{T}_0(1+z)^{1-f_{GA}(z)}$ finding a similar $\chi^2$ with Eq.~(\ref{eq:reconst}). We want to clarify that although the GA provide a smooth and differentiable function at all redshifts they are a non-parametric algorithm, hence the traditional statistical comparison based on Bayesian inference is ambiguous. Then the use of quantitative criterion such as the Bayesian Information Criterion (BIC), Akaike Information Criterion (AIC) or Evidence Ratio cannot be used in this case to make a fair and consistent comparison. For this reason we compare the best-fit $\chi^2$ for the GA and $\Lambda\text{CDM}$ as can be seen in Table~(\ref{tab:chi2}).\\

\begin{figure}[!t]
\centering
\includegraphics[width=0.7\textwidth]{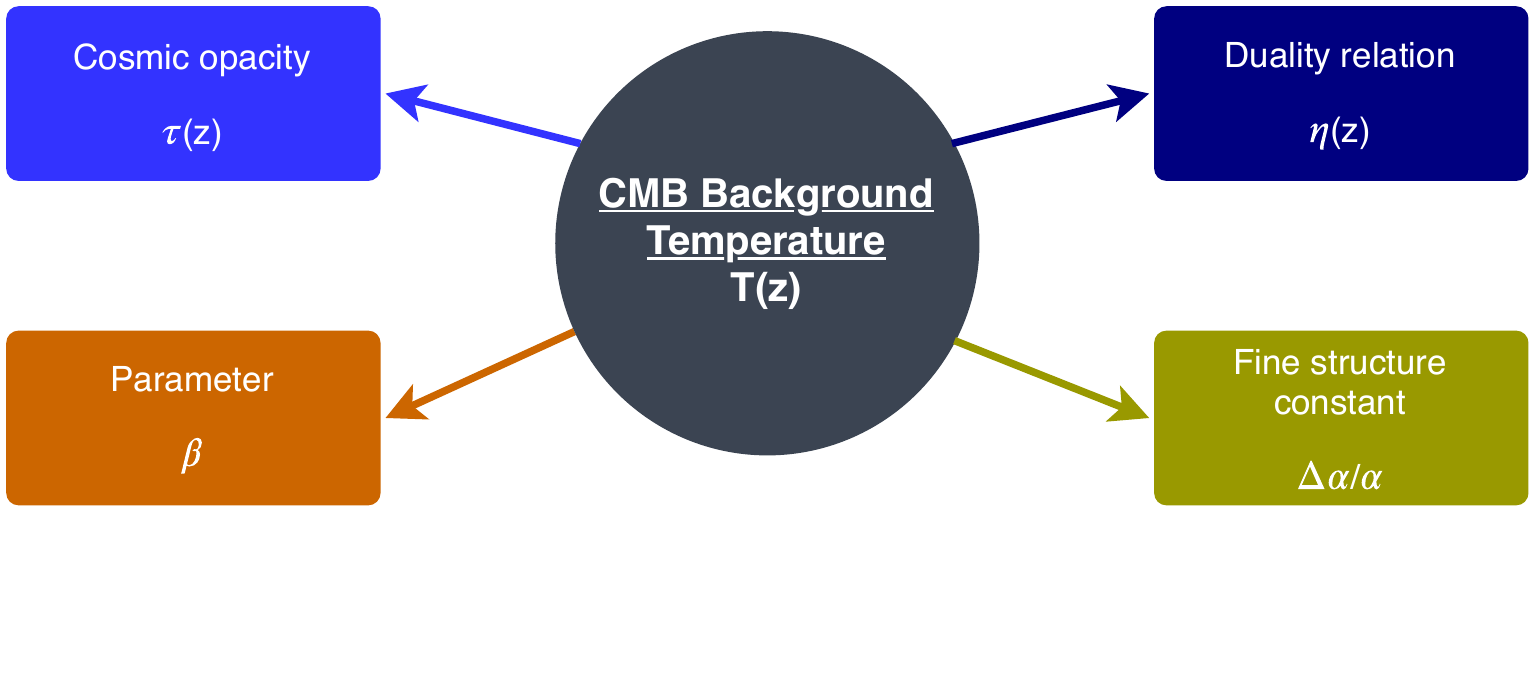}
\caption{A flowchart of the reconstructed functions using our Machine Learning algorithm for the T(z) data.  \label{fig:flowchart}}
\end{figure}

\begin{table}[!t]
\caption{The $\chi^2$ for \lcdm and GA using the $T(z)$ data.\label{tab:chi2}}
\begin{centering}
\begin{tabular}{ccc}
  & $T(z)$  \\\hline
$\chi^2_{\Lambda \text{CDM}}$  & 29.176 \\\hline
$\chi^2_{GA}$  & 28.816 \\\hline
\end{tabular}
\par
\end{centering}
\end{table}
Following, we present some theoretical context for the remaining derived quantities we will reconstruct and in Fig.~\ref{fig:flowchart} we present a flowchart of our fitting process for illustration purposes. We stress that the robustness of the GA approach resides in the fact that is a non-parametric method and does not require an initial prior or a cosmological fiducial model, obtaining constraints in a model independent approach. We also make a reminder that these constraints are not independent of each other since all of them parameterize potential deviations from the background temperature relation of the CMB.

\begin{figure*}[!t]
\centering
%\hspace*{-4mm}
\includegraphics[width = 0.49\textwidth]{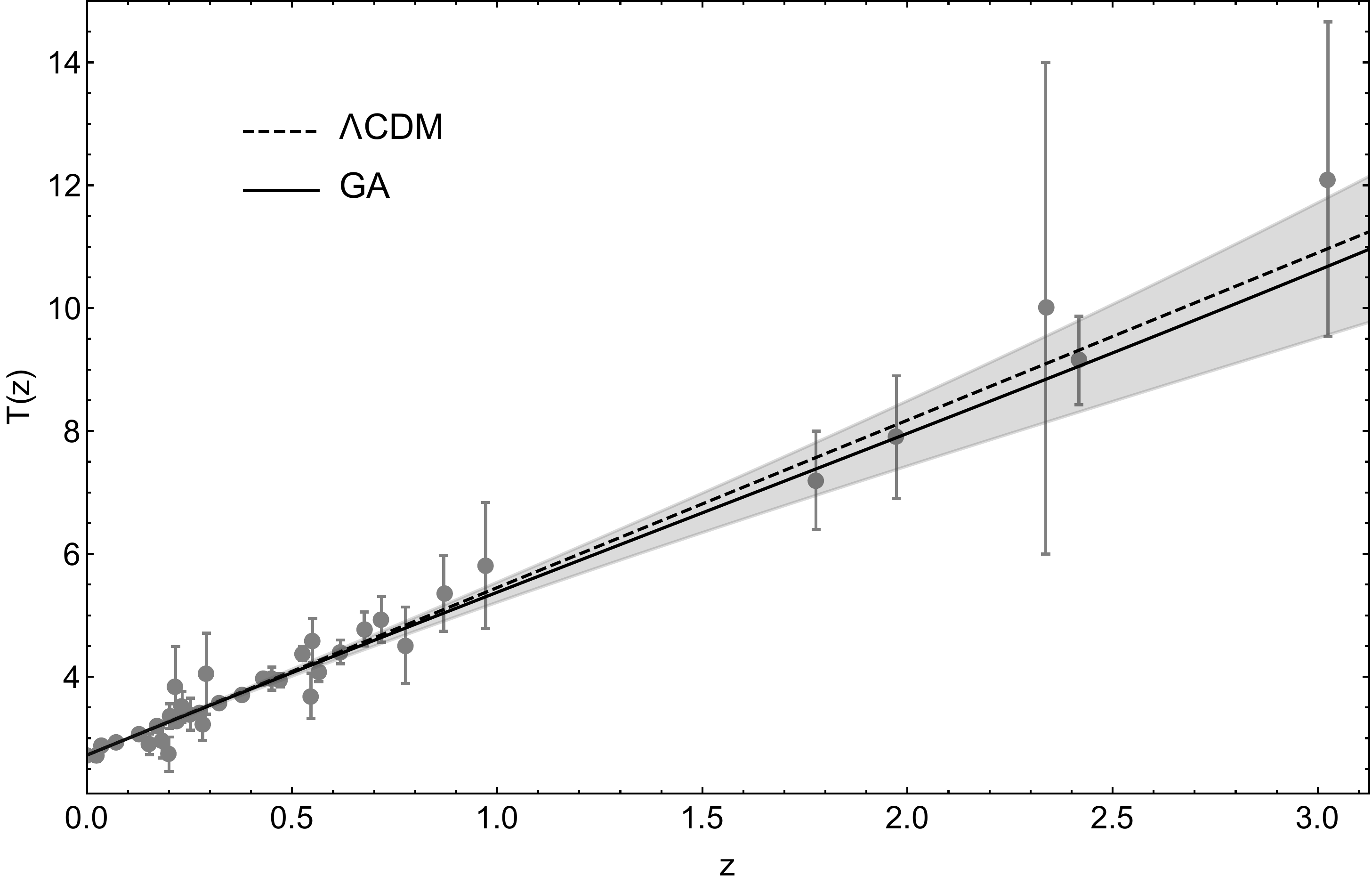}
\includegraphics[width = 0.49\textwidth]{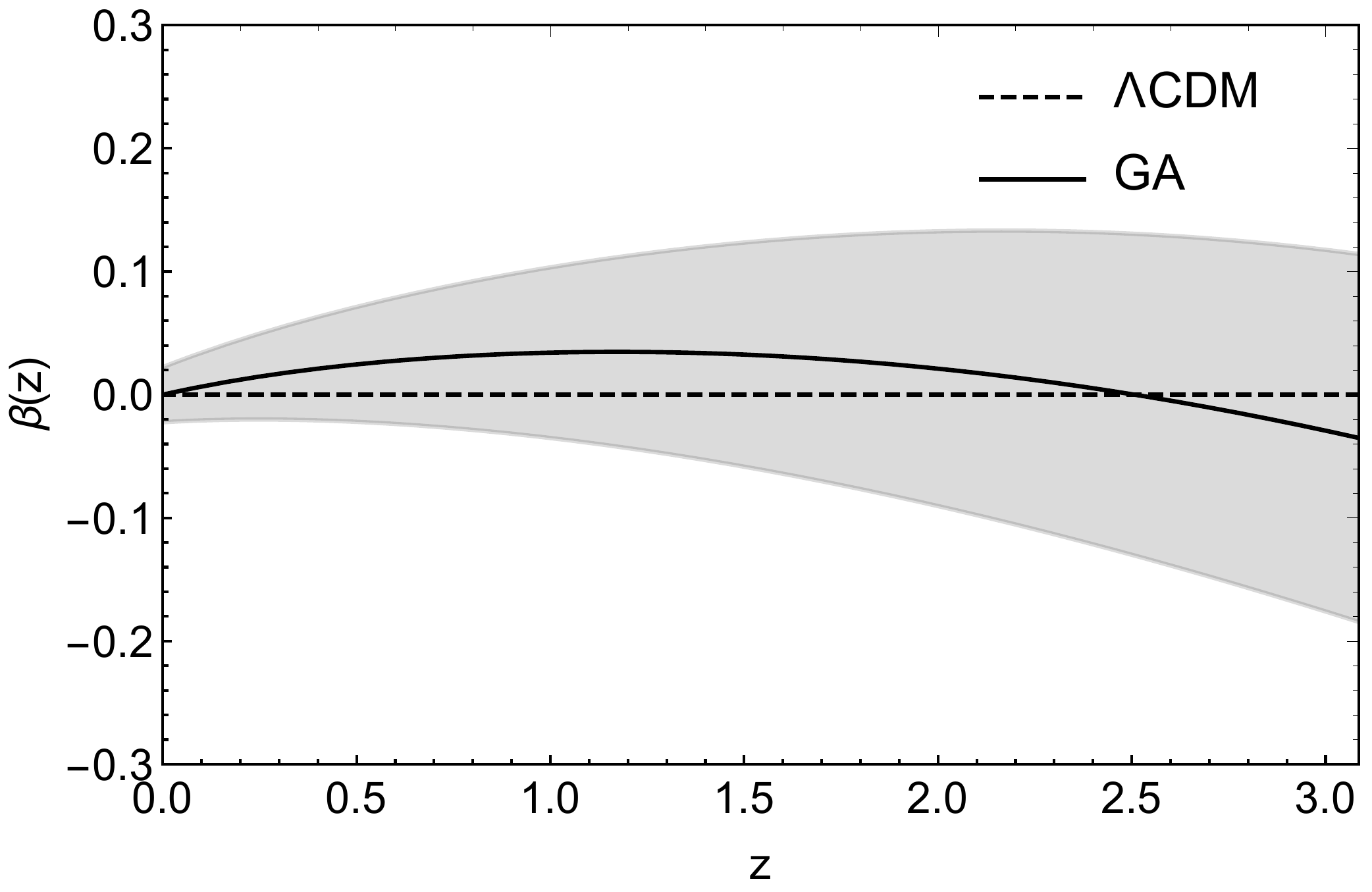}
\caption{Left: The $\text{T}(z)$ data compilation shown as grey points along with the \lcdm best-fit (dashed line) and the GA best-fit (solid black line). The shaded gray regions corresponds to the $1\sigma$ errors of the GA. Right: The reconstruction of the $\beta$ parameter. Both cases are consistent with the \lcdm model. \label{fig:Data}}
%\vspace*{-1mm}
\end{figure*}

\subsection{$\beta$ parameter}
If we assume that the expansion of the Universe is adiabatic, then the hot Big Bang model predicts that the CMB temperature evolves proportional to $(1+z)$. One can parameterize possible deviations to this linear law as
\begin{equation}\label{eq:beta}
T(z)=T_0(1+z)^{1-\beta},
\end{equation}
where $\beta$ is a parameter that would account for adiabatic photon production $\beta>0$ or destruction $\beta<0$. This phenomena can occur for example in decaying dark energy models \cite{Jetzer:2011kw} where DE interacts with matter by the creation of photons, affecting in turn the CMB spectrum \cite{Muller:2012kv}. From Eq.~(\ref{eq:beta}) we find that we can write $\beta$ as the logarithmic derivative of $\text{T}(z)$
\begin{equation}\label{eq:beta1}
  \beta(z)=1-(1+z)\frac{d \ln \left(T(z)/T_0 \right)}{dz},
\end{equation}
where $\text{T}(z)$ represents our best-fit reconstructed function, Eq.~(\ref{eq:reconst}), and $\text{T}_0$ our derived parameter, Eq.~(\ref{eq:H0bf}). In the right panel of Fig.~\ref{fig:Data} we present our reconstruction of the $\beta$ parameter where the dashed line is the prediction from \lcdm  and the GA best-fit is the solid black line. The shaded gray regions corresponds to the $1\sigma$ errors of the GA. We find that our model independent approach is consistent with an adiabatic universe and the conservation of photon number.

\subsection{Duality relation and the cosmic opacity}

%%%%%%%%%%%%%%%%%%%%%%%%%%%%%%%%%%%%%%%%%%%%%%%%%%%%%%%%%%
The distance duality relation (DDR) defines a connection between the luminosity distance $d_L$ and the angular diameter distance $d_A$ in the following way
\begin{equation}\label{eq:ddrr}
\eta(z)\equiv \frac{d_L}{(1+z)^2d_A}=1,
\end{equation}
where any deviations from $\eta(z) \neq 1$ would be a hint for new physics, e.g. that the Universe is opaque. This relation is valid under the condition of the conservation of photon number in cosmic evolution, gravity must be described by a metric theory and the travel of photons along null geodesics \cite{ellis2007definition} holding true for all curved space-times. The DDR has been tested from different datasets ranging from radio galaxies and ultra compact radio sources \cite{Bassett:2003vu}, The CMB \cite{Rasanen:2015kca,Lazkoz:2007cc,Ellis:2013cu}, Baryon Acoustic Oscillations (BAO) \cite{Ma:2016bjt,cardone2012testing}, $H$ $21$cm signal from disk galaxies \cite{Khedekar:2011gf}, Gamma Ray Bursts \cite{Holanda:2014lna} and high redshift quasars \cite{Zheng:2020fth} among others. If we assume the temperature-redshift relation of the CMB from the \lcdm model, i.e. $\text{T}(z)=\text{T}_0(1+z)$, then the factor $1+z$ can be written as $1+z=\text{T}(z)/\text{T}_0$. Inserting this relation into the rhs of Eq.~(\ref{eq:ddrr}) we find that
\begin{equation}\label{eq:dlt}
  \frac{d_L}{d_A}=(1+z)^2=\left(\frac{T(z)}{T_0}\right)^2.
\end{equation}
Substituting the ratio $\frac{\text{d}_\text{L}}{\text{d}_\text{A}}=\left(\frac{\text{T}(z)}{\text{T}_0}\right)^2$ from Eq.~(\ref{eq:dlt}) in Eq.~(\ref{eq:ddrr}) we see that the DDR can also be written in terms of the redshift temperature relation of the CMB as
\begin{equation}\label{eq:ddrt}
  \left(\frac{T(z)}{T_0(1+z)}\right)^2\equiv \eta(z),
\end{equation}
as it is also shown in Ref.~\cite{Rana:2015feb} and which should be equal to unity in the \lcdm model. The above relation is directly connected to the cosmic opacity $\tau(z)$ as we will show below and therefore we can use our GA reconstruction on $\eta(z)$ and $\tau(z)$ to constrain the transparency of the universe.

If we have an opaque universe, the photon flux collected by the observers is lowered by a factor $e^{-\tau(z)}$, and the observed luminosity distance $d_{L,obs}$ can be expressed as \cite{Zhou:2019gda}
\begin{equation}\label{eq:tau}
  d_{L,obs}(z)=d_{L,true}(z)e^{\tau(z)/2},
\end{equation}
where $\tau(z)$ denotes the opacity parameter between an observer at $z = 0$ and a source at $z$, and physically it gives us information about how transparent is the universe or in other words it connotes the optical depth associated to the cosmic absorption. This parameter can mimic a dark energy behaviour \cite{Holanda:2014lna} and reconstructions for the parameter $\tau(z)$ have been done in the past \cite{more2009cosmic,avgoustidis2009consistency,nair2012cosmic} and recently it has been tested from Gravitational Waves mock data from the third generation of the Einstein Telescope and using Gaussian Processes \cite{Zhou:2019gda}. From Eq.~(\ref{eq:ddrr}) and Eq.~(\ref{eq:tau}) we can see that
\begin{equation}
  e^{\tau(z)/2}=\eta(z),
\end{equation}
then using Eq.~(\ref{eq:ddrt}) we see that $\tau(z)$ and our reconstruction for $\text{T}(z)$ are connected in the following way
\begin{equation}\label{eq:taure}
  \tau(z)=4\ln \left(\frac{T(z)/T_0}{(1+z)}\right).
\end{equation}

\subsection{Fine structure constant}

Fundamental constants, which we assume to be constant over space-time, are described operationally, meaning that nature does not force it to be constant. They are not given by the theory and must be obtained experimentally. For a review on the variation of fundamental constants see \cite{Landau:2020vkr}. Here we will probe the interesting case where the fine structure constant $\alpha=\frac{e^2}{\hbar c}$ is not invariant and we will express its relative  variation as $\Delta \alpha/\alpha$. If there are eventually signatures of a variation it would have imprints in different physical mechanisms such as the CMB anisotropies \cite{Uzan:2010pm}. Constraints on this variation, both temporal and spatial, have been performed already \cite{Clara:2020efx,deMartino:2016bjx,deMartino:2016tbu,Hees:2014lfa,Colaco:2020ndf,Lopez-Honorez:2020lno,Wilczynska:2020rxx}, and this variation can be produced for example through an evolving scalar field which is coupled to the electromagnetic Lagrangian \cite{deMartino:2016bjx} producing violations in the photon number conservation. This will give rise to both, variations of the fine structure constant and violations of the standard $T_{CMB}(z)$ law. A class of models where this occurs is the Bekenstein-Sanvik-Barrow-Magueijo (BSBM) model \cite{Sandvik:2001rv}, where the electric charge is allowed to vary. Although such theories preserve the local gauge and Lorentz invariance, the fine structure constant will vary during the matter dominated era. The corresponding action is
\begin{equation}
S=\int d^{4} x \sqrt{-g}\left(\mathcal{L}_{g}+\mathcal{L}_{m}-\frac{\omega}{2} \partial_{\mu} \psi \partial^{\mu} \psi-e^{-2 \psi} \mathcal{L}_{e m}\right),
\end{equation}
where $\mathcal{L}_{g}=\frac{1}{16\pi G}R$ is the Hilbert-Einstein Lagrangian plus the matter fields $\mathcal{L}_{m}$, the third term is the kinetic term for the scalar field $\psi$ and, finally, the last term couples the scalar field with the standard electromagnetic Lagrangian $\mathcal{L}_{e m}=\frac{F^{\mu \nu} F_{\mu \nu}}{4}$. Then the governing evolution equation of the radiation energy reads
\begin{equation}\dot{\rho}_{\gamma}+4 H \rho_{\gamma}=2 \dot{\psi} \rho_{\gamma},\end{equation}
with
\begin{equation}
  \frac{\alpha}{\alpha_0}=\exp^{2\left(\psi-\psi_0\right)}.
\end{equation}
For this type of models, assuming adiabaticity, the relation between the evolution of the CMB temperature $\text{T}(z)$ and the variation of the fine structure constant is expressed as \cite{Avgoustidis:2013bqa}
\begin{equation}\label{eq:alphy1}
  T(z)/T_0\sim(1+z)\left(1+\frac{1}{4} \frac{\Delta \alpha}{\alpha}\right).
\end{equation}
Since $\Delta \alpha/ \alpha$ is expected to be small experimentally, a more general phenomenological relation that can be tested observationally and can be seen as a good approximation for a wider range of couplings is expressed as \cite{Clara:2020efx}
\begin{equation}\label{eq:alphy}
  T(z)/T_0\sim(1+z)\left(1+\varepsilon \frac{\Delta \alpha}{\alpha}\right),
\end{equation}
or instead
\begin{equation}
  \frac{\Delta T_{CMB}}{T}=\frac{T_{CMB}(z)-T_{CMB,std}(z)}{T_{CMB,std}(z)}\sim \varepsilon \frac{\Delta \alpha}{\alpha},
\end{equation}
where $T_{CMB,std}(z)$ represents the evolution of the CMB temperature in the $\Lambda$CDM model, i.e $T_{CMB,std}(z)=T_0(1+z)$. The coefficient $\varepsilon$ depends on the specific model under consideration and it is commonly assumed to be of order unity \cite{deMartino:2016tbu}, hence we will consider as a test case $\varepsilon=1$. Writing $\frac{\Delta \alpha}{\alpha}$ as a function of $\text{T}(z)$ we find
\begin{equation}\label{eq:fine}
  \frac{\Delta \alpha}{\alpha}\sim-\frac{(1+z)-T(z)/T_0}{\varepsilon\left(1+z\right)}.
\end{equation}
This relation Eq.~(\ref{eq:fine}), can be tested for both, time and/or spatial variations of $\alpha$ \cite{Avgoustidis:2013bqa}. To test for spatial variations of the fine structure, we would have to resolve the CMB temperature at the cluster location using multi-frequency measurements of the thermal Sunyaev Zel'dovich effect (TSZ). This has already been performed in \cite{deMartino:2016tbu,deMartino:2016bjx} and we plan to do a similar study using the Genetic Algorithms (GA) but such analysis is left for a future work. In this paper we will limit to constrain how the fine structure constant $\alpha$ changes with distance, or in other words, test its temporal evolution using the CMB temperature at redshift between 0 and 3, with the data compilation found in Table~\ref{tab:data} which comes from SZ observations at low redshifts and from observations of spectral lines at high redshift. To the best of our knowledge, this is the first time that temporal variations on the fine-structure constant $\frac{\Delta \alpha}{\alpha}$ are constrained in a model independent and non parametric approach using the GA.
%In our results we will compare two cases, a simplistic adiabatic limit $\varepsilon=1/4$ and a limit case $\varepsilon=1$.

\section{Results \label{sec:results}}
\begin{figure*}[!t]
\centering
%\hspace*{-4mm}
\includegraphics[width = 0.49\textwidth]{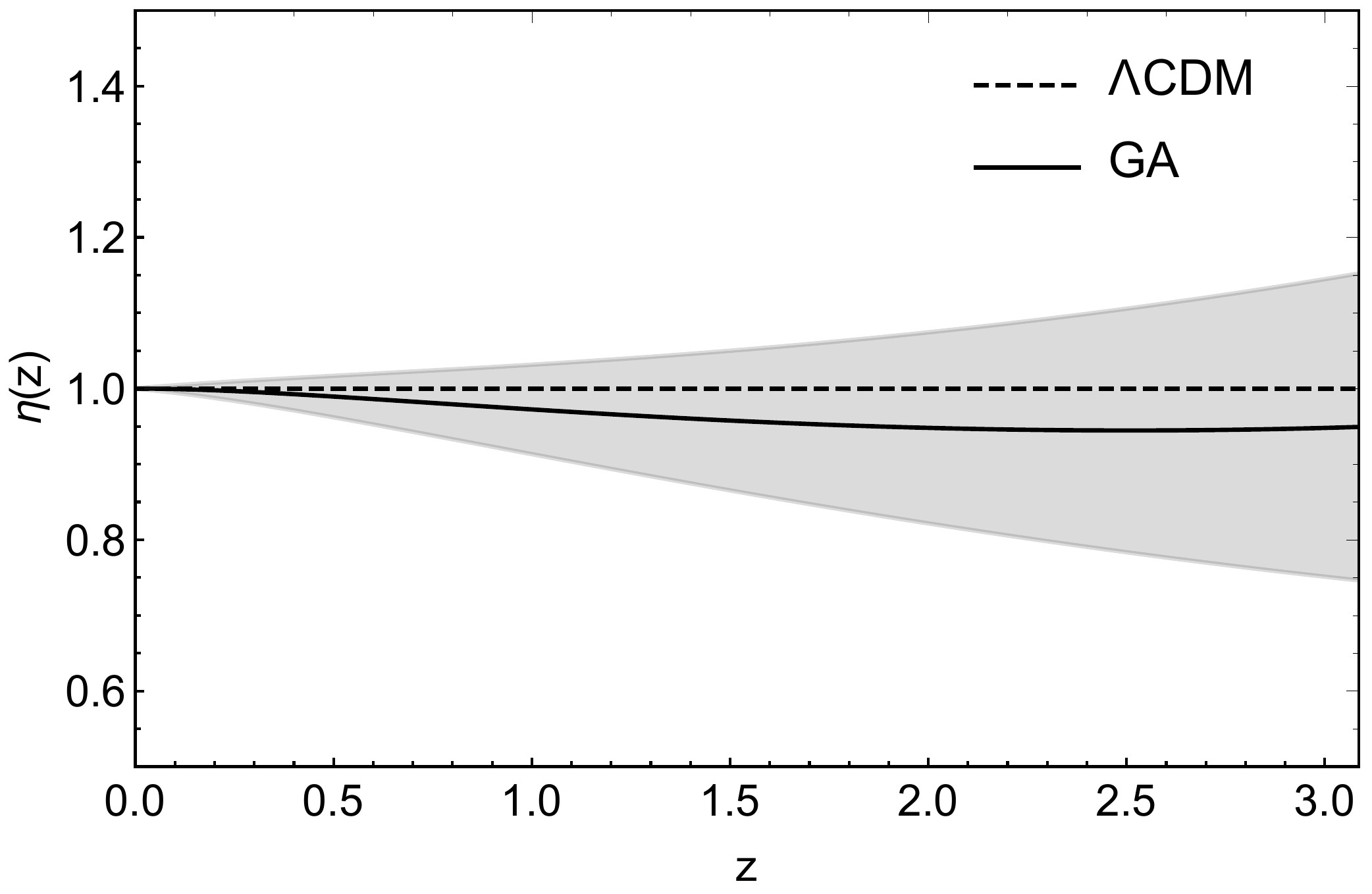}
\includegraphics[width = 0.49\textwidth]{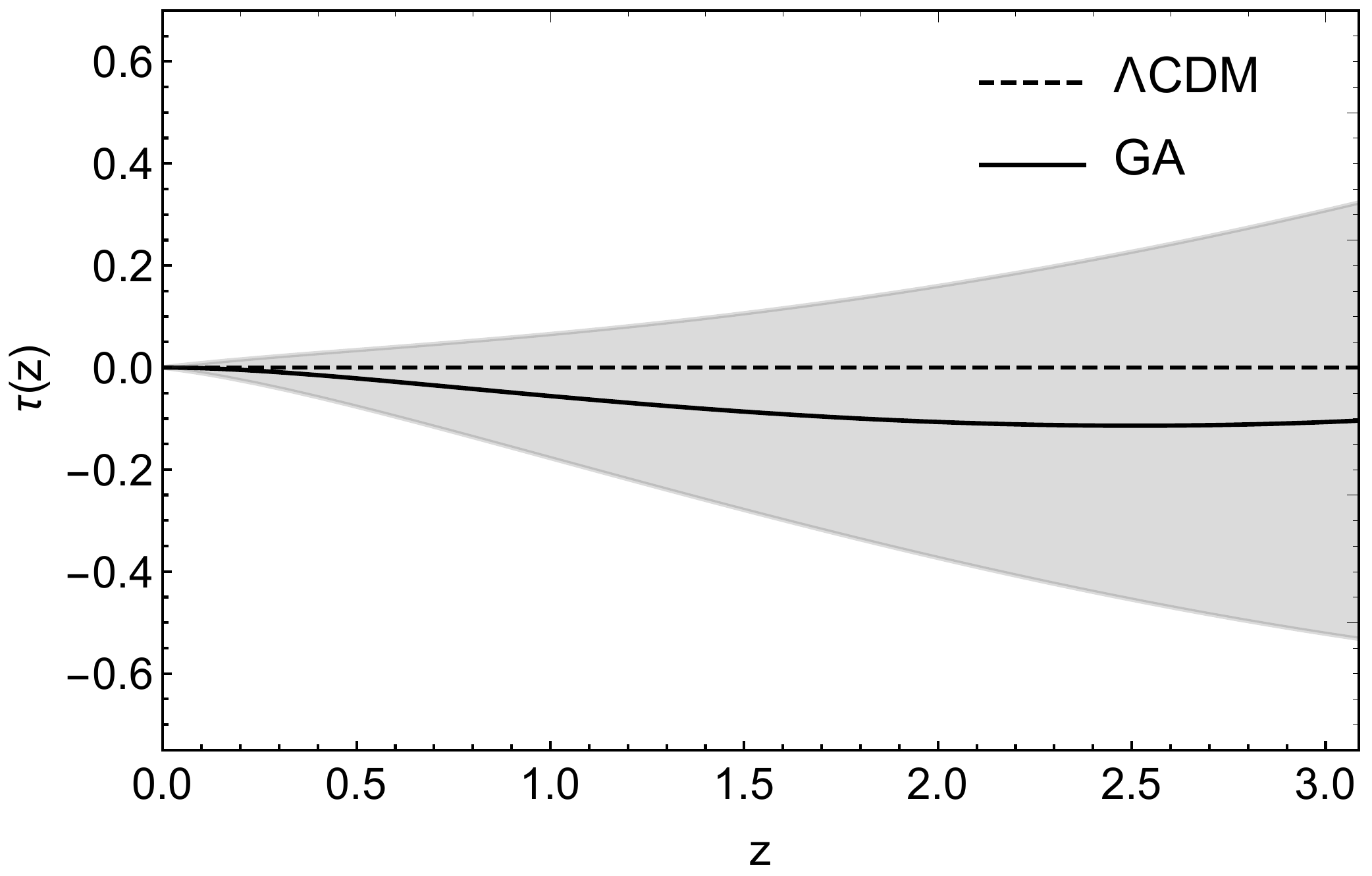}
\caption{Left: The reconstruction of the duality relation $\eta(z)$. Right: The reconstruction of the cosmic opacity $\tau(z)$ . In both cases the \lcdm best-fit corresponds to the dashed line and the GA best-fit to the solid black line along with the 1$\sigma$ errors (gray regions). Both cases are consistent with the \lcdm model.  \label{fig:dual_opacity}}
%\vspace*{-1mm}
\end{figure*}

\begin{figure*}[!t]
\centering
%\hspace*{-4mm}
\includegraphics[width = 0.7\textwidth]{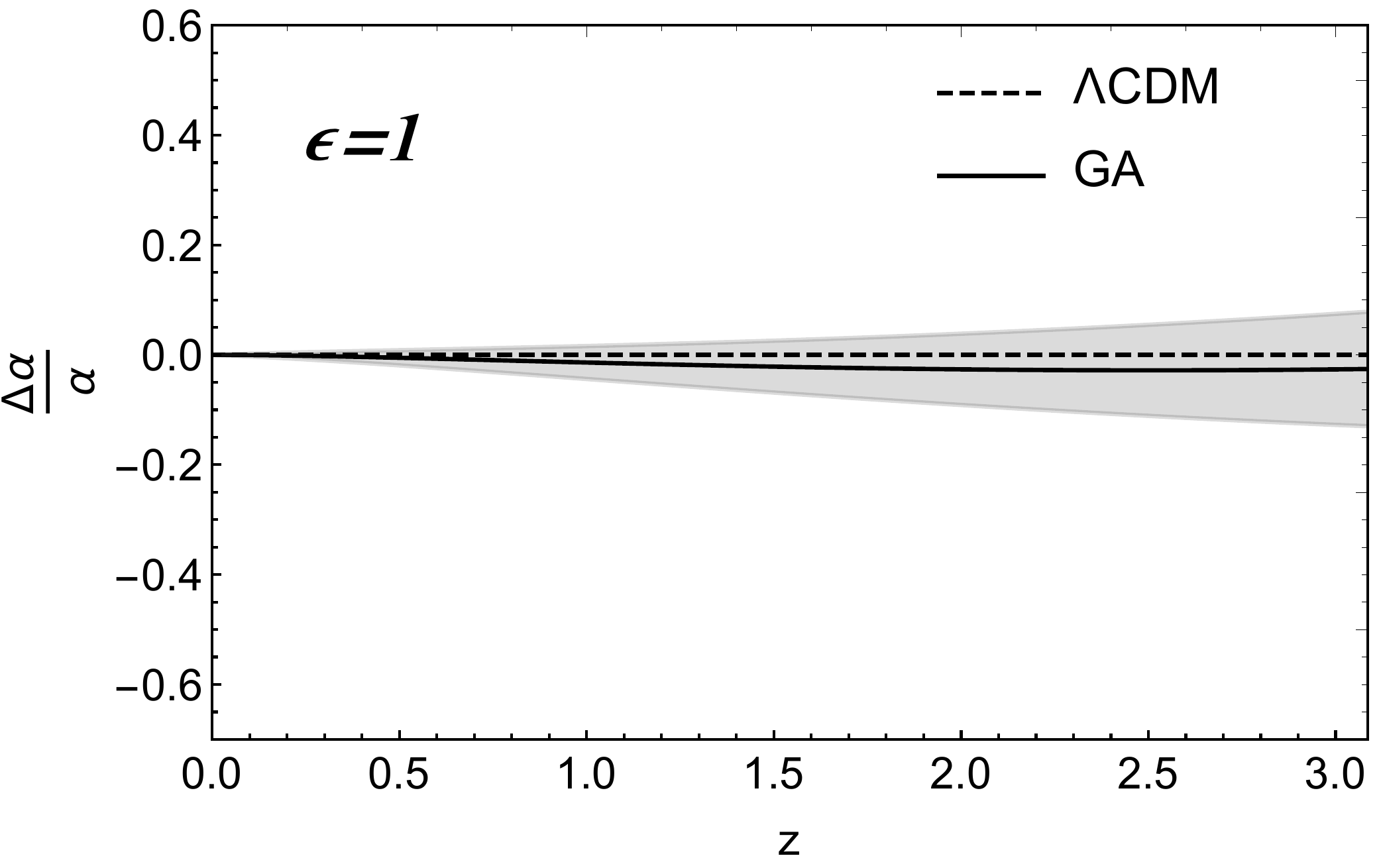}
\caption{The reconstruction of the temporal variation of the fine structure constant $\frac{\Delta \alpha}{\alpha}$ for $\varepsilon=1$. The dashed line corresponds to a non varying constant and the GA best-fit to the solid black line along with the 1$\sigma$ errors (gray regions). We found that our reconstruction is consistent with a fine structure constant that is not temporally varying.  \label{fig:beta_alph}}
%\vspace*{-1mm}
\end{figure*}

In this Section we present our best-fit reconstructions for the parameter $\beta$, the duality relation $\eta(z)$, the opacity parameter $\tau(z)$ and temporal variations on the fine structure constant $\frac{\Delta \alpha}{\alpha}$. Inserting our reconstructed $\text{T}(z)$ function Eq.~(\ref{eq:reconst}) in Eqs.~(\ref{eq:beta1}),(\ref{eq:ddrt}),(\ref{eq:taure}) and (\ref{eq:fine}) we derived the following constraints at redshift $z=0$
\ba
\hspace{-0.5cm}\beta(z=0) &=&0.0000\pm 0.0224, \\
\hspace{-0.5cm}\eta(z=0) &=&1.0000\pm 0.0002 \hspace{0.1cm}, \\
\hspace{-0.5cm}\tau(z=0) &=&0.0000\pm 0.0004, \\
%\hspace{-0.5cm}\frac{\Delta \alpha}{\alpha}(z=0,\varepsilon=1/4) &=&0.0000\pm 0.0004, \\
\hspace{-0.5cm}\frac{\Delta \alpha}{\alpha}(z=0,\varepsilon=1) &=&0.0000\pm 0.0001,
\ea
and at redshift $z=3.025$ where we have our last data point
\ba
\hspace{-0.5cm}\beta(z=3.025) &=&-0.0309\pm 0.1475, \\
\hspace{-0.5cm}\eta(z=3.025) &=&0.9483\pm 0.1986 \hspace{0.1cm}, \\
\hspace{-0.5cm}\tau(z=3.025) &=&-0.1062\pm 0.4188, \\
%\hspace{-0.5cm}\frac{\Delta \alpha}{\alpha}(z=3.025,\varepsilon=1/4) &=&-0.1048\pm 0.4078, \\
\hspace{-0.5cm}\frac{\Delta \alpha}{\alpha}(z=3.025,\varepsilon=1) &=&-0.0262\pm 0.1020.
\ea
From the numbers given above, we can see that both at low and high redshifts our constraints are consistent with the \lcdm model.\\
In the left and right panel of Fig.~\ref{fig:dual_opacity} we show the reconstruction of the duality relation $\eta(z)$ and the cosmic opacity parameter $\tau(z)$ respectively. In both cases the expected value from \lcdm corresponds to the dashed line and the GA best-fit to the solid black line along with the 1$\sigma$ errors (gray regions). Both cases are consistent with photon number conservation and a transparent universe and hence with the \lcdm model. Finally, in Fig.~\ref{fig:beta_alph} we show our reconstruction of the variation of the fine structure constant for $\varepsilon=1$ which is consistent with a non varying constant within the $1\sigma$ region. Overall, for all our reconstructions we find no evidence of deviations within the $1\sigma$ region from the well established $\Lambda\text{CDM}$ model.

%Both cases are consistent in being constant and not varying and we have checked that as the value of $\varepsilon$ increases, the errors decrease.\\

%Although we find for all our reconstructions very good agreement with the \lcdm model specially at low redshifts, we appreciate a general trend for negative values at high redshifts for the parameter $\beta$, the cosmic opacity parameter $\tau(z)$ and variations on the fine structure constant $\frac{\Delta \alpha}{\alpha}$ and values of $\eta(z)$ below $1$, thus suggesting the existence of a mechanism that violates the conservation of the photon number.

%\begin{figure*}[!t]
%\centering
%\hspace*{-4mm}
%\includegraphics[width = 0.49\textwidth]{alpha14.pdf}
%\includegraphics[width = 0.49\textwidth]{alpha1.pdf}
%\caption{Left: The reconstruction of the variations of the fine structure constant $\frac{\Delta \alpha}{\alpha}$ for $\varepsilon=1/4$. Right: The reconstruction of the variations of the fine structure constant $\frac{\Delta \alpha}{\alpha}$ for $\varepsilon=1$. In both cases the \lcdm best-fit corresponds to the dashed line and the GA best-fit to the solid black line along with the 1$\sigma$ errors (gray regions). Both cases are consistent with the \lcdm model and we see that as the value of $\varepsilon$ increases, the errors decrease.  \label{fig:beta_alph}}
%\vspace*{-1mm}
%\end{figure*}

\section{Conclusions \label{sec:conclusions}}
We have presented a model independent and non-parametric reconstruction of data coming from the redshift evolution of the CMB temperature which spans over a redshift range of $0\le z \le 3.025$ with a Machine Learning algorithm without assuming any dark energy model, an adiabatic universe or photon number conservation. In particular we used the genetic algorithms which avoids the dependency on an initial prior or a cosmological fiducial model. From our reconstruction we have provided constraints and updated estimates in a novel approach on the $\beta$ parameter from the parametrisation $\text{T}(z)=\text{T}_0(1+z)^{1-\beta}$, the duality relation $\eta(z)$ and the cosmic opacity parameter $\tau(z)$. Furthermore we place constraints on a temporal varying fine structure constant $\alpha$, which would have signatures in a broad spectrum of physical phenomena such as the CMB anisotropies. It is important to notice that our constraints are not independent of each other since all of them parameterize in diverse ways potential deviations from the temperature-redshift relation. Within uncertainties, our model independent approach is consistent with the standard view of $\text{T} \propto \left(1+z\right)$ having found no strong discrepancies within the $1\sigma$ region with the \lcdm model. Finally, our results evidence that a transparent universe is preferred at $1\sigma$.\\

%Finally, we have seen that our reconstructions seem to have a preference for violations of the photon number conservation at high redshifts, suggesting the presence of a mechanism that produces a net photon destruction.\\

%%%%%%%%%%%%%%%%%%%

\textbf{Numerical Analysis File}: The Genetic Algorithm code used by the author in the analysis of the paper will be released upon publication of the paper at the GitHub repository \href{https://github.com/RubenArjona}{https://github.com/RubenArjona}.

\acknowledgments
It is a pleasure to thank S. Nesseris for useful discussions and the anonymous referees for helpful remarks. We also acknowledge support from the Research Projects FPA2015-68048-03-3P [MINECO-FEDER], PGC2018-094773-B-C32 and the Centro de Excelencia Severo Ochoa Program SEV-2016-0597.

\appendix
\section{Fisher matrix approach \label{sec:fisher}}
\begin{figure}[!t]
\centering
\includegraphics[width=0.75\textwidth]{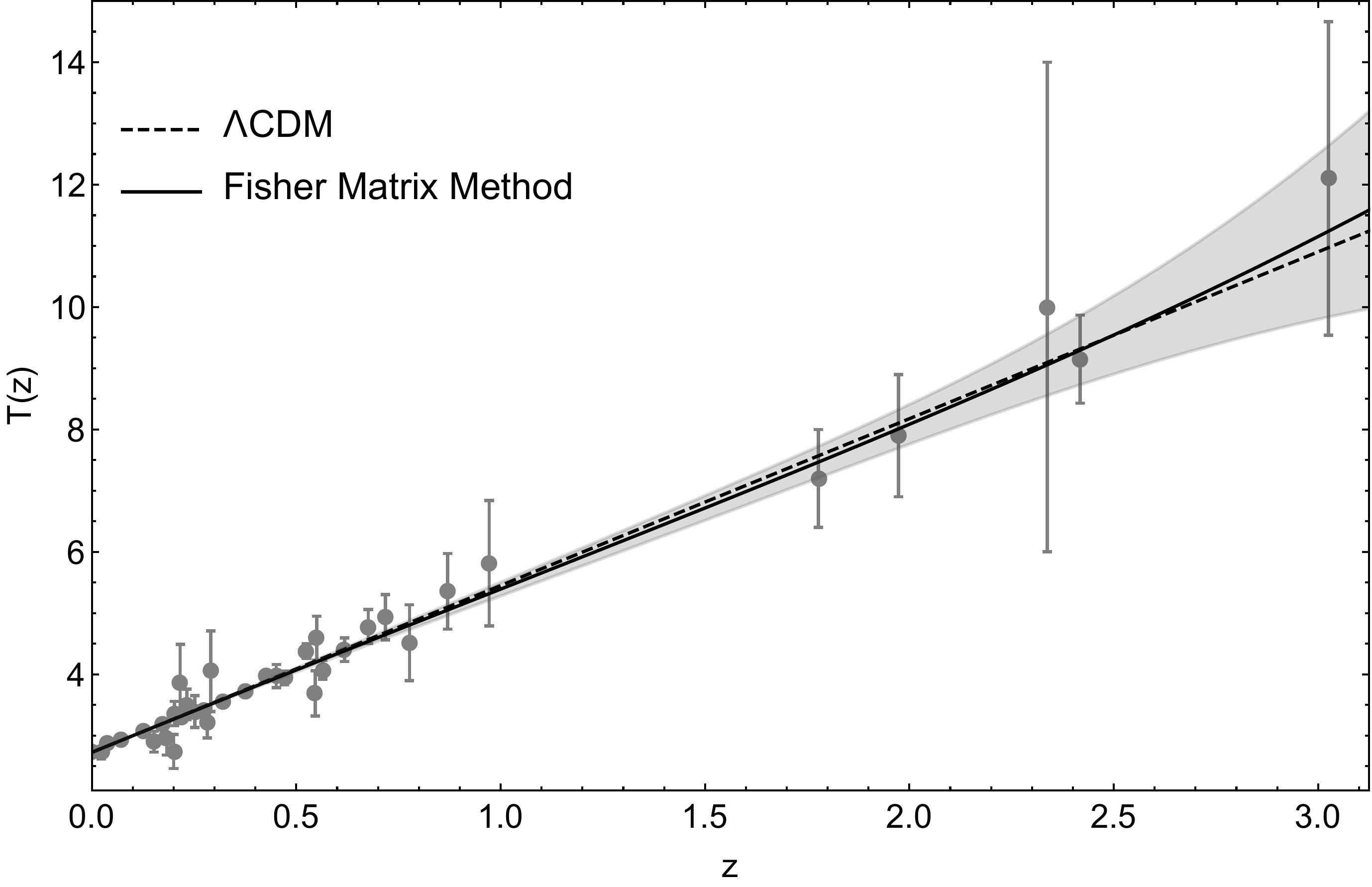}
\caption{The $\text{T}(z)$ data compilation shown as grey points along with the \lcdm best-fit (dashed line) and the Fisher matrix method best-fit (solid black line). The shaded gray regions corresponds to the $1\sigma$ errors of the Fisher matrix method and is consistent with the GA approach, see Fig.~\ref{fig:Data}. \label{fig:Fisher}}
\end{figure}

To evaluate the rigor of the path integral approach for the error analysis of the GA we compare it numerically with the Fisher matrix approach. We chose the following function which could be used to test deviations from the $\Lambda\text{CDM}$ model
\begin{equation}\label{eq:funct}
  f(z;a,b)=T_0(1+z)^{1+ax+bx^2},
\end{equation}
where $z$ is the redshift and $a$ and $b$ are constant numbers. Then we fitted the model $f(z;a,b)$ of Eq (\ref{eq:funct}) by minimizing the $\chi^2$
\begin{equation}\label{eq:chi2}
  \chi^2(a,b)=\underset{i}{\sum}\left(\frac{y_i-f(z_i;a,b)}{\sigma_{y_i}}\right)^2.
\end{equation}
The best-fit value is given by $\left(a,b\right)_{min}=\left(a=-0.0264\pm 0.0502,b=0.0106\pm 0.0231\right)$ with a $\chi^2_{min}=28.892$. The shaded gray region from Fig.~\ref{fig:Fisher} is the $1\sigma$ error following a Fisher Matrix approach \cite{Nesseris:2012tt}. The error of our best-fitted function $f(z;a,b)$ is obtained from
\begin{equation}
  \sigma_f(z)^2=\underset{i,j}\sum C_{ij}\partial_i f(z;a,b)\partial_j f(z;a,b)\mid_{min},
\end{equation}
which is evaluated at the best fit \cite{press2007numerical} and the dummy variables $(i,j)$ correspond to our parameters $(a,b)$. The covariance matrix $C_{ij}$ is obtained from the inverse of the Fisher matrix $C_{ij}=F_{ij}^{-1}$ where
\begin{equation}\label{fisher}
  F_{ij}=\frac{1}{2}\partial_{ij}\chi^2(a,b)\mid_{min},
\end{equation}
evaluated at the best-fit. Comparing the shaded gray regions from the Fisher matrix method see Fig.~\ref{fig:Fisher} and the GA approach, see Fig.~\ref{fig:Data} we see that the path integral approach \cite{Nesseris:2012tt,Nesseris:2013bia} is robust.

\section{Data compilation and error analysis \label{sec:errors}}
In our analysis we use 37 points and the compilation can be found in Table~\ref{tab:data}. The main advantage of our compilation is that it spans over a wide redshift range of $0\le z \le 3.025$, thus testing the $\beta$ parameter, the duality relation $\eta(z)$, the cosmic opacity parameter $\tau(z)$ and temporal variations on the fine structure constant $\frac{\Delta \alpha}{\alpha}$ up to high redshifts. The compilation of the data is assumed to be uncorrelated since there is no public access to any correlation matrix.\\
The background temperature of the CMB can be measured at both high and low redshifts. For the former, it can be recovered through fine-structure transitions of atomic or molecular species in cool absorption-line systems along the line of sight to high redshift quasars \cite{LoSecco:2001zz}. For low redshifts it can be obtained from Sunyaev-Zel'dovich (SZ) effect in clusters of galaxies. The existing measurements at high redshifts still have large error bars and the majority of the points can be only treated as upper limits. However, in the near future, with high resolution spectroscopy with larger telescopes, the precision of these measurements can be competitive with local interstellar data. For the last data point of  Table~\ref{tab:data} we compute the error as
\begin{equation}\label{dataerr}
  \sigma_T(z=3.025)=\sqrt{\frac{\sigma_u^2+\sigma_d^2}{2}}=2.562,
\end{equation}
where $\sigma_i$ is the error on the temperature estimates and $N$ is the number of the observational data used.

\begin{table}[!t]
\caption{Compilation of the CMB temperature-redshift relation $\text{T}(z)$ measurements used in this analysis and related references. We used 37 data points.	 \label{tab:data}}
\centering
\begin{tabular}{ccccccccc}
\\
\hline\hline
$z$  & $\text{T(K)}$ & Ref.   \\
\hline
0.000    & 2.72548 $\pm$ 0.00057  & \cite{Fixsen:2009ug}   \\
0.023    & 2.72 $\pm$ 0.10  & \cite{luzzi2009redshift}   \\
0.152    & 2.90 $\pm$ 0.17  & \cite{luzzi2009redshift}   \\
0.183    & 2.95 $\pm$ 0.27  & \cite{luzzi2009redshift}   \\
0.200    & 2.74 $\pm$ 0.28  & \cite{luzzi2009redshift}   \\
0.202    & 3.36 $\pm$ 0.20  & \cite{luzzi2009redshift}   \\
0.216    & 3.85 $\pm$ 0.64  & \cite{luzzi2009redshift}   \\
0.232    & 3.51 $\pm$ 0.25  & \cite{luzzi2009redshift}   \\
0.252    & 3.39 $\pm$ 0.26  & \cite{luzzi2009redshift}   \\
0.282    & 3.22 $\pm$ 0.26  & \cite{luzzi2009redshift}   \\
0.291    & 4.05 $\pm$ 0.66  & \cite{luzzi2009redshift}   \\
0.451    & 3.97 $\pm$ 0.19  & \cite{luzzi2009redshift}   \\
0.546    & 3.69 $\pm$ 0.37 & \cite{luzzi2009redshift}   \\
0.550    & 4.59 $\pm$ 0.36  & \cite{luzzi2009redshift}   \\
2.418    & 9.15 $\pm$ 0.72  & \cite{srianand2008first}   \\
1.777    & 7.20 $\pm$ 0.80  & \cite{cui2005molecular}   \\
1.973    & 7.9 $\pm$ 1  & \cite{Ge:1996it}   \\
2.338    & 10 $\pm$ 4  & \cite{Srianand:2000wu}   \\
0.037    & 2.888 $\pm$ 0.039  & \cite{Hurier:2013ona}   \\
\hline\hline
\end{tabular}
\begin{tabular}{ccccccccc}
\\
\hline\hline
$z$  & $\text{T(K)}$ & Ref.   \\
\hline
0.072    & 2.931 $\pm$ 0.017  & \cite{Hurier:2013ona}   \\
0.125    & 3.059 $\pm$ 0.032  & \cite{Hurier:2013ona}   \\
0.171    & 3.197 $\pm$ 0.030  & \cite{Hurier:2013ona}   \\
0.220    & 3.288 $\pm$ 0.032  & \cite{Hurier:2013ona}   \\
0.273    & 3.416 $\pm$ 0.038  & \cite{Hurier:2013ona}   \\
0.322    & 3.562 $\pm$ 0.050  & \cite{Hurier:2013ona}   \\
0.377    & 3.717 $\pm$ 0.063  & \cite{Hurier:2013ona}   \\
0.428    & 3.971 $\pm$ 0.071  & \cite{Hurier:2013ona}   \\
0.471    & 3.943 $\pm$ 0.112  & \cite{Hurier:2013ona}   \\
0.525    & 4.380 $\pm$ 0.119  & \cite{Hurier:2013ona}   \\
0.565    & 4.075 $\pm$ 0.156  & \cite{Hurier:2013ona}   \\
0.618    & 4.404 $\pm$ 0.194  & \cite{Hurier:2013ona}   \\
0.676    & 4.779 $\pm$ 0.278  & \cite{Hurier:2013ona}   \\
0.718    & 4.933 $\pm$ 0.371  & \cite{Hurier:2013ona}   \\
0.777    & 4.515 $\pm$ 0.621  & \cite{Hurier:2013ona}   \\
0.870    & 5.356 $\pm$ 0.617  & \cite{Hurier:2013ona}   \\
0.972    & 5.813 $\pm$ 1.025  & \cite{Hurier:2013ona}   \\
3.025    & $12.1^{+1.7}_{-3.2}$ & \cite{Molaro:2001jv}   \\
    &  &   \\
\hline\hline
\end{tabular}
\end{table}
The $1\sigma$ errors of $\beta$, $\eta(z)$, $\tau(z)$ and $\frac{\Delta \alpha}{\alpha}$ were computed following the traditional error propagation, since it has been shown \cite{Arjona:2020kco} that is in agreement with the errors obtained using the definition of the standard deviation $\delta g^2=\langle g^2\rangle-\langle g\rangle^2$, where $g$ is a quantity formed by a function $f$. For the $\beta$ parameter, since it is defined as
\begin{equation}
  \beta(z)=1-(1+z)\frac{d \ln \left(T(z)/T_0 \right)}{dz},
\end{equation}
following the aforementioned approach we find that the error of $\beta$, e.g. $\delta \beta$ is
\begin{equation}\label{eq:betas}
  \delta \beta(z)=-(1+z)\frac{d\left(\delta T(z)/T(z)\right)}{dz},
\end{equation}
where $\text{T}(z)$ is our best-fit function given by the GA and its $1\sigma$ error obtained through the path integral approach is $\delta \text{T}(z)$. Similarly we can derive the rest of the errors. For the duality relation $\eta(z)$ we have that
\begin{equation}
 \eta(z)=\left(\frac{T(z)/T_0}{(1+z)}\right)^2,
\end{equation}
then
\begin{equation}
  \delta  \eta(z)=\frac{2T(z)\delta T(z)}{T_0^2(1+z)^2}.
\end{equation}
For the cosmic opacity parameter defined as
\begin{equation}
  \tau(z)=4\ln \left(\frac{T(z)/T_0}{(1+z)}\right),
\end{equation}
we have
\begin{equation}
  \delta \tau(z)=\frac{4}{T(z)/T_0}\ln \left(\delta T(z)/T_0\right),
\end{equation}
and finally for the variation of the fine structure constant
\begin{equation}
  \frac{\Delta \alpha}{\alpha}=-\frac{(1+z)-T(z)/T_0}{\varepsilon\left(1+z\right)},
\end{equation}
we found the following
\begin{equation}
  \delta \left(\frac{\Delta \alpha}{\alpha}\right)=\frac{1}{(1+z)\varepsilon}\frac{\delta T(z)}{T_0}.
\end{equation}

\section{The Genetic Algorithms \label{sec:GA}}
In this section we provide a brief background to our implementation. Genetic Algorithms (GA), which are a particular Machine Learning method and performs unsupervised symbolic regression of data, applies an evolutionary programming algorithm which aims to find a solution by the process of simulated evolution, allowing to explore many possibilities, using the biological theory of genetics and the principle of survival of the fittest \cite{suleman1997genetic}.\\
The algorithm starts with a random initial population of individuals (solutions), generated for the problem of study, and these then go through evolution by means of reproduction, crossover and mutation of individuals until some ending criteria is achieved, for example when it is obtained a solution with fitness greater than some predefined threshold or when it is reached a maximum number of generations. This last condition assures that the code finishes, even if one doesn't obtain the desired level of fitness. The GA require that only the parameters or the grammar for the problem is specified. Thereafter the algorithm applied to search for a solution is mostly problem-independent and is able to reconstruct an analytic function that fits the data. In our analysis we consider the T(z) data shown in Table ~\ref{tab:data}. Our predefined grammar consisted on an orthogonal basis of functions: exp, log and polynomials and a set of operations $+,-,\times,\div$. We fixed the size of the initial population to 2000 and the maximum number of generations to 1000. It has been shown that the choice of the grammar only affects the rate of convergence of the GA \cite{Bogdanos:2009ib}.\\
A selection, known as the population, of potential solutions is kept throughout the life cycle of the GA. At the beginning of the algorithm, one can specify any needed physical assumptions, for example in our T(z) reconstruction we imposed that the value of the CMB temperature relation today is $T(z=0)=T_0$ and that all functions found by the GA are differentiable and continuous without no singularities in the redshift range covered by the data. In this way we can avoid overfitting and unphysical functions. Then a population of solutions is generated and the fitness of each member is estimated by a $\chi^2$ statistic, using as input the T(z) data. This population is then modified through the mechanisms of evolution to result eventually in individuals that are closer to the solution than these initial random ones. This process is repeated thousands of times and with different random seeds to properly explore the functional space.\\
To be more specific, we will briefly explain the principles of evolution of reproduction, crossover and mutation. The mechanism of all evolutionary change in the GA is reproduction. The reproduction operation grants the population to progress from one generation into the next. This progression happens in the most natural way possible, favouring the individuals with a better fit. Two popular methods are the roulette wheel selection and tournament selection. In the former approach, which is the one implemented in the analysis, the fitnesses of all individuals in the population are ordered into a list and then summed. Within the range of the sum, a random number is generated and the fitnesses in the list are summed again until the random number
is obtained or exceeded. The last individual in the list is the one picked. In the tournament selection, random small groups of two individuals are chosen from the population for crossover and mutation and their fitnesses are compared, picking the dominant member of each group. See Ref.~\cite{Bogdanos:2009ib} for more details. After the reproduction selection, the best fitting functions in each generation are selected and go through the stochastic operations of crossover and mutation. The operation of crossover consists on the merger of two or more different individuals, producing two new individuals (children), who might be better than their parents. This is comparable to genetic crossover as observed in living organisms. In our implementation we set a probability between $60\%\sim 75\%$ to produce crossover. Finally, the operation of mutation dwells with a random modification in an individual. During reproduction, fitter individuals in a population are selected more regularly than others. This leads to some individuals not being selected into the next generation. These are generally the least fit individuals. However, they may accommodate within their structure parts of a better solution. This genetic material would be forgotten since the individuals are no longer propagated. In order to recover from this loss, the operation of mutation allows to change the genetic material randomly. The mutation rate was chosen to have a probability of the order of $10\%\sim 25\%$.\\
The evaluation of the errors for our reconstruction is computed analytically through a path integral approach, defined previously in \cite{Nesseris:2012tt,Nesseris:2013bia}, which scans over all possible functions that might be produced by the GA. This method has been extensively contrasted with the Fisher Matrix approach and the Bootstrap Monte Carlo \cite{Nesseris:2012tt}. This assessment of the errors is necessary for a proper statistical interpretation of the data.\\

\bibliographystyle{JHEP}
\bibliography{Tz_data}
\end{document}